\documentclass{aa} % for a paper on 1 column  
\usepackage{mathabx}
\usepackage{txfonts}
\usepackage{graphicx}   % Including figure files
\usepackage{amsmath}    % Advanced maths commands
\usepackage{amssymb}    % Extra maths symbols
\usepackage{multirow}   % Extra command \multirow
\usepackage{array}      % 
\usepackage{mathtools}
\usepackage{xcolor} 
\usepackage[none]{hyphenat}
\usepackage{makecell}
\usepackage{hyperref}

\hypersetup{colorlinks=true,linkcolor=blue,citecolor=blue,filecolor=blue,urlcolor=blue,}
\makeatletter
\renewcommand*\aa@pageof{, page \thepage{} of \pageref*{LastPage}}
\makeatother

\definecolor{green_comm}{RGB}{0,160,0}
\definecolor{orange}{RGB}{255,165,0}
\graphicspath{./Figures/}

\newcommand{\fxuv}{F_{\rm XUV}}

\newcommand{\LNF}{L_{\rm XUV}^{\rm  Q}}

\newcommand{\LXUV}{L_{\rm XUV}}

\newcommand{\Mlost}{M}
\newcommand{\MlostQ}{M^{Q}}
\newcommand{\MlostF}{M^{F}}
\newcommand{\RE}{$R_{\Earth}~$}
\newcommand{\ME}{$M_{\Earth}~$}
\newcolumntype{P}[1]{>{\centering\arraybackslash}p{#1}}
\let\oldpropto\propto
\renewcommand{\propto}{\, \oldpropto \,}
\begin{document} 
\title{Why M-dwarf flares have limited impact on the atmospheric evaporation of sub-Neptunes and Earth-sized planets}
\titlerunning{The role of flares in exoplanet photoevaporation}
\authorrunning{Caldiroli et al.}
\author{Andrea Caldiroli\inst{1} 
    \and Francesco Haardt\inst{1,2,3}
    \and Elena Gallo,\inst{4}\\
         George King\inst{4}
    \and Juliette Becker\inst{5}
    \and Federico Biassoni\inst{1,2}
    \and Riccardo Spinelli\inst{6,1}    
}
\institute{
    % 1
    Como Lake Center for Astrophysics (CLAP), DiSAT, Universit\`a degli Studi dell'Insubria, via Valleggio 11, I-22100 Como, Italy
\and
    % 2
    INAF -- Osservatorio Astronomico di Brera, Via E. Bianchi 46, I-23807 Merate, Italy 
\and
    % 3
    INFN, Sezione Milano-Bicocca,P.za della Scienza 3, I-20126 Milano, Italy
\and
    % 4
    Department of Astronomy, University of Michigan, 1085 S University, Ann Arbor, Michigan 48109, USA
\and 
    %5
    Department of Astronomy, University of Wisconsin-Madison, 475 N. Charter St., Madison, WI 53706, USA
\and
    % 5
    INAF – Osservatorio Astronomico di Palermo, Piazza del Parlamento 1, I-90134, Palermo, Italy
}
\date{Received; accepted}
\abstract{
M-type stars are prime targets for exoplanet searches within their habitable zones (HZs). These stars also exhibit significant magnetic flaring activity, particularly during their first billion years, which can potentially accelerate the evaporation of the hydrogen-helium envelopes of close-in planets. We employ the time-dependent photoionization hydrodynamics code \texttt{ATES} to investigate the impact of flares on atmospheric escape, focusing on an Earth-sized and a sub-Neptune-sized planet orbiting an early M-type star at distances of 0.01, 0.1, and 0.18--0.36~AU—the inner and outer edges of the HZ. Stellar flaring is modeled as a 1 Gyr-long high-activity phase followed by a 4 Gyr-long low-activity phase, each characterized by an appropriate flare frequency distribution. We find that flares have a modest impact--less than a factor of two--on the cumulative atmospheric mass loss, with the greatest absolute enhancement occurring when the planets are at their closest separation. However, the relative  enhancement in mass loss between flaring and non-flaring cases is greater at larger orbital separations. This trend arises because, as stellar irradiation fluctuates between quiescent levels and peak flares, the proportion of time that a planet spends in the energy-limited versus recombination-limited mass loss regimes depends on its orbital separation. Additionally, we demonstrate the existence of a characteristic flare energy, intermediate between the minimum and maximum values, that maximizes the fractional contribution to flare-driven mass loss. Our results indicate that the flaring activity of M-dwarfs does not significantly affect the atmospheric retention of close-in planets, including those within the HZ. The potential occurrence of rare super-flares, which current observational campaigns may be biased against, does not alter our conclusions.
}
\maketitle
%
%---------------------------------------------------------------------------- %
%
\section{Introduction}
\label{sec:intro}
%
%---------------------------------------------------------------------------- %
%
The habitable zones (HZs) of M dwarfs are at smaller orbital separations than for Sun-like stars. Combined with the lower mass and radius of their stars, this leads to potentially habitable planets being far more easily detectable around M dwarfs too, and thus they have been the a primary focus of many recent detection efforts. However, M dwarfs are also known to flare more frequently \citep[e.g.][]{Hawley2014,Youngblood2017,Tilley2019}, and so understanding the impact of these flares on the evolution and chemistry of planets around them is even more important than it is for Sun-like stars.

Our understanding of stellar flares, particularly at a statistical level, has blossomed over the past decade or two. One of the main driving forces of this has been the advent of long-baseline, high-precision photometric surveys like \textit{Kepler}, \textit{K2}, and \textit{TESS}, which have led to numerous studies of the optical properties of stellar flares across a range of stellar types, ages etc. \citep[e.g.][]{Davenport2014,Hawley2014,Davenport2016,Doyle2018,Gunther2020,Pietras2022,Feinstein2024}. 

Stellar X-ray and extreme-ultraviolet (EUV; together XUV) emission can drive substantial escape from planetary atmospheres \citep[e.g.][]{Lammer2003}, in some cases removing the entire primordial H/He envelope \cite[e.g.][]{Owen2012}. At a population level, such atmospheric stripping is one of the leading mechanisms proposed to explain the observed Neptunian desert \citep{Owen2018} and radius valley \citep{Owen2017}, both of which are dearths of planets in radius-period space which cannot be explained by selection effects. In addition to their heightened flare rates, M dwarfs can remain at a ``saturated'' activity level (often characterized in terms of the ratio of the XUV, or just X-ray, and bolometric luminosities, $L_{\rm XUV}/L_{\rm bol}$, with the saturated level being $\approx 10^{-3}$) for as long as 1-2\,Gyr \citep{Engle2024}, as opposed to the $\sim 100$\,Myr of Sun-like stars \citep{Jackson2012}.

While only typically a small portion of the bolometric flare emission, flaring events at XUV wavelengths often outshine the rest of the quiescent coronal emission, albeit only for a short period of time. Often for simplicity reasons, many lifetime photoevaporation modeling efforts solely consider the quiescent emission though \citep[e.g.][]{Lopez2012,Luger2015,Chen2016,Owen2017,Malsky_2023,King2024}, and neglect XUV flaring. Another possible factor is our relatively poor understanding of stellar XUV flares on a statistical level, at least compared to the optical.

In general, we anticipate and have observed that many of the relationships applicable to optical stellar flares also qualitatively apply to the XUV spectrum. For instance, younger stars tend to flare more frequently due to their higher magnetic activity. 
Some divergence in terms of quantitative results in the XUV would perhaps not be surprising though given these wavelengths are known to trace a different component of the flare emission compared to optical, as demonstrated by the well-known Neupert effect \citep{Neupert1968}. Gyro-synchrotron emission in the radio and hard X-ray ($\gtrsim 10$\,keV, and their thermal proxy which peaks in the near-UV around the U-band) is associated with the early, impulsive stage, whereas the soft XUV traces the heating of plasma by the non-thermal electrons. 
Detections of this effect for other stars dates back three decades \citep{Hawley1995}, and a typical manifestation of this is soft X-ray flares being time-lagged and longer in duration compared to NUV or optical flares \citep[e.g.][]{Guedel2002}. Multi-wavelength detections of the same flares also continue to sometimes yield unexpected results, such as optically quiet stars with regular FUV flaring \citep{Jackman2024}.

Only a few studies have looked in detail at X-ray flare properties for larger ensembles of events. \citet{Pye2015} studied about 130 events from 70 stars observed with \textit{XMM-Newton} ranging in type from F to M. More recently, \citet{Zhao2024} examined a similar number of flares from specifically Solar-like stars across a range of energies with \textit{Chandra}, notably finding a flare frequency distribution (FFD) similar to other wavelengths, suggesting the proportion of the total flare emission in X-rays could be roughly constant. \citet{Getman2021} have also looked at much larger sample of over a thousand flares from pre-main sequence stars, finding some consistency in the FFD with optical and X-ray observations of their older counterparts. This small but growing body of works therefore suggest that, to first-order, X-ray FFDs could be similar to other wavelengths, even if the temporal properties of individual flares (e.g., their durations) can be divergent.

The impact of stellar flares on atmospheric losses in exoplanets is thought to be an important factor in shaping planetary evolution and habitability, especially for close-in planets orbiting active stars (\citealt{ketzer2023}, and references therein). They may also affect the observed transit absorption line profiles \citep[e.g.,][]{Zhilkin2024} and alter the chemistry of secondary atmospheres on Earth-like planets \citep{Segura2010, Tilley2019}. 

In terms of connecting XUV flaring up with its contribution to atmospheric escape, a slew of studies have modeled the response of an atmosphere to individual flares \citep[for a review, see Section 4.3 of][]{Hazra2025}. 
For a Solar flare, \citet{Lee2018} inferred a 20\% increase in the escape rate of oxygen from Mars, largely driven by EUV wavelengths. Many of the flares we observe from other stars are several orders of magnitude more powerful than those we observe on the Sun. \citet{Bisikalo2018} modeled the response of HD\,209458b's atmosphere to flares up to 1000 times brighter in XUV compared to quiescence, finding significant increases of in some cases over an order of magnitude in the escape rate. While also finding an escape enhancement, \citet{Chadney2017} found that the effect of increased XUV irradiation was not enough to explain observed temporal changes in the Ly-$\alpha$ transit which had been suggested to result from an observed flare some hours previous \citep{LDE2012}. Related, a recent 2D modeling study by \citet{Gillet2025} also found that stellar flare timescales are generally too short to have a large impact on the atmospheric mass loss and associated Ly-$\alpha$ transit signals. \citet{Chadney2017} however did find that a coronal mass ejection associated with the flare could explain the deeper transit, with a more recent study by \citet{Hazra2022} reaching similar conclusions. For smaller exoplanets, \citet{France2020} determined a significant enhancement of the possible escape rate in flare for planets around older ($\sim 10$\,Gyr) M dwarfs such as Barnard's star. Their results suggested that the flare duty cycle could be of great importance for atmospheric stability, and therefore habitability, at these older ages. Perhaps unsurprisingly, flares also affect the chemical processes in the upper atmosphere across a range of sizes \citep[e.g.][]{Louca2023}.

On a lifetime level, there have also been some recent studies incorporating repeated flaring into the modeling of escape-driven planetary evolution. Using an energy-limited approach, which invariably maximizes mass loss, \citet{Atri2021} determined that for most stars the quiescent XUV irradiation is the dominant driver of escape from terrestrial planets, with only a minor contribution from flares, but that this contribution can be as high as 20\% for planets around mid-to-late M dwarfs. Focusing on M dwarfs, \citet{doAmaral2022} found that flares could double the water loss from Earth-like planets across the first Gyr of their lives. In a more recent study, \citet{doAmaral2025} examined the young early M star AU Mic, finding that its planet d will likely be stripped of its entire atmosphere over the next few Myr, in a process that can be driven by the quiescent XUV irradiation alone. However, the study also concludes that any planets further away in the habitable zone of AU Mic could be much more significantly shaped by flares, particularly at later times, in line with \citet{France2020}. These studies model the impact of flares by assuming that the stellar quiescent luminosity is increased by an average factor, which varies depending on the chosen frequency distribution of the flares and the stellar activity phase. To the best of our knowledge, no time-dependent investigation exists in the literature that addresses the effects of repeated flares on atmospheric escape. 

In this work, we use the time-dependent, photo-ionization hydrodynamics code ATmospheric EScape \citep[ATES,][]{Caldiroli_2021} to examine the effect of stellar XUV flares on atmospheric escape, and assess the lifetime-cumulative impact of these flares on planetary evolution. We describe our models and methodologies in Section \ref{sec:method}. We present our results in Section \ref{sec:res}, and place them into further context with some discussion in Section \ref{sec:discussion}. 
%
%---------------------------------------------------------------------------- %
%
\begin{figure}
    \centering
    \includegraphics[width=.9\columnwidth]{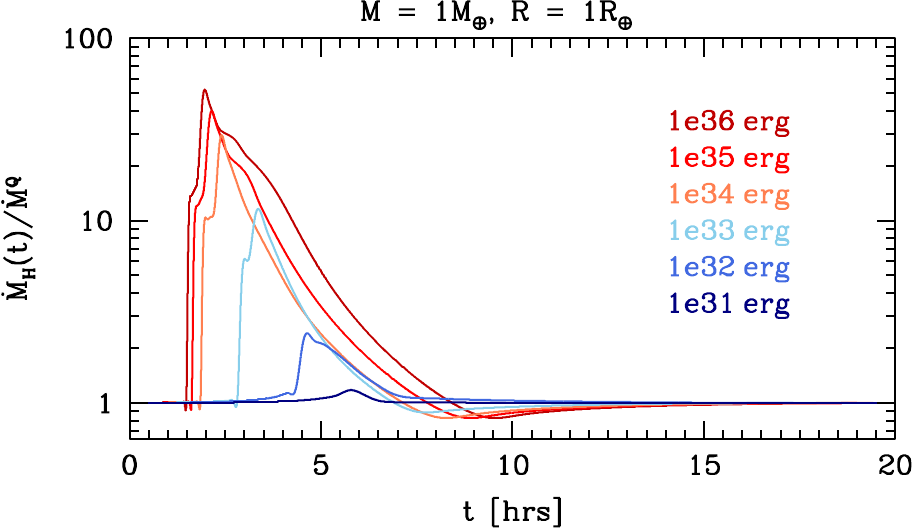}\\[10pt]
    \includegraphics[width=.9\columnwidth]{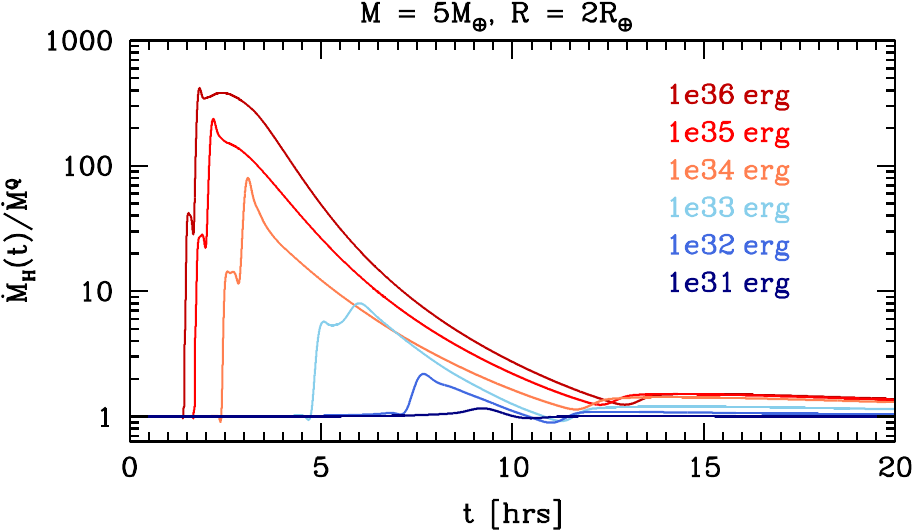}
    \caption{Temporal evolution of atmospheric mass loss in response to stellar flares of different energies. The ratio of the instantaneous mass loss rate to the quiescent mass loss rate is shown for input flares ranging from $10^{31}$~erg (dark blue) to $10^{36}$~erg (dark red). Values are provided at the system's Hill radius. The top and bottom panels refer to the Earth- and sub-Neptune-sized planet, respectively.} 
    \label{fig:t_Mdot}
\end{figure}
\section{Methodology}
\label{sec:method}
%
%---------------------------------------------------------------------------- %
%
\subsection{Flare-driven mass loss contrast}
\label{subsec:Flareprofile}

In order to asses the impact of stellar flares on atmospheric escape we model the time-dependent stellar XUV ($10 - 912$\r{A}) luminosity as:
\begin{equation}
    \LXUV(t) = \left[ 1 + \delta_F(t) \right] \LNF,
    \label{eq:Ltime}
\end{equation}
where $\delta_F(t)$ represents the luminosity increase due to a single flaring event, commonly referred to as flare contrast, while $\LNF$ denotes the baseline (quiescent) stellar luminosity. Each flare has a characteristic duration, $\Delta T$, and an energy in the XUV range of \citep{Kowalski2024rev}: 
\begin{equation}
    E = \LNF \int_{\Delta T}{\delta_F(t)\,dt}.
    \label{eq:flareEnergy}
\end{equation}

The flare time evolution is modeled with a linear rise and an exponential decline. We assume a flare rise time of 500 s and an exponential decay timescale of 1,000 s, independent of the flare energy. These rise and decay timescales are chosen to be representative of the mean XUV flare, while acknowledging that timescales are typically longer at shorter wavelengths. This is because the UV acts as a thermal counterpart proxy for the gyro-synchrotron emission of the initial flare impulse itself, whereas the thermal soft X-ray emission is produced by the surrounding coronal plasma heated by the reconnection event (see \citealt{Kowalski2024}, and references therein). For the purposes of this investigation, we disregard potential spectral hardening during the flare \citep{Osten2015, Pye2015, Kowalski2024}.\par
The atmospheric mass that is lost from a planet due to the XUV irradiation from the host star can be expressed as:
\begin{equation}
    \Mlost = \MlostQ \left( 1 + \frac{\Delta\MlostF}{\MlostQ} \right),
\end{equation}
where $\MlostQ$ represents the mass lost due to photoevaporation driven by the (constant) quiescent irradiation, while $\Delta\MlostF$ is the additional mass lost due to flares. 

The fractional mass-loss enhancement can be written as 
%%%%%%%%%%%%%%%%%%%%%%%%%%%%%
\begin{equation}
 \frac{\Delta\MlostF}{\MlostQ} = \int_{E_{\rm min}}^{E_{\rm max}}  dE\,\frac{d \dot N}{dE}\,\,\int_{\Delta T}{dt\,\left(\frac{\dot M_E^F(t)}{\dot M^Q}-1\right)},
 \label{eq:DMflare}
\end{equation}
%%%%%%%%%%%%%%%%%%%%%%%%%%%%%
where $\dot M^F_E$ is the time dependent mass loss rate occurring during a flare of energy $E$, and $\dot M^Q$ is the constant mass loss rate in the absence of flares. The number of flares per unit energy per unit time (e.g., per day, or Flare Frequency Distribution FFD), $d\dot N/dE$, is typically modeled as a power law of the form $d\dot N/dE \propto E^{-\alpha}$ \citep{Lacy1976, Audard2000}. The normalization and exponent of the FFD are derived from fits to observational data and may vary between different stellar types, energy bands, etc. \citep[e.g.,][]{Davenport2019, Feinstein2024}.
Similarly, the range over which the energy integral in Eq.~\ref{eq:DMflare} extends depends on the star's age and activity level.

From Eq.~\ref{eq:DMflare} we can define the contribution to the fractional mass-loss enhancement from a single flare of energy $E$ as 
\begin{equation}
    \delta_M(E)=\frac{1}{\Delta T} \int_{\Delta T}{dt\,\left(\frac{\dot M_E^F(t)}{\dot M^Q}-1\right)}.
    \label{eq:DMsingleflare}
\end{equation}
Note that, while the fractional mass-loss enhancement in Eq.~\ref{eq:DMflare} is in principle independent upon the actual flare durations $\Delta T$, $\delta_M$ is not. Given the characteristic exponential decay of flare luminosity as a function of time, in evaluating Eq.~\ref{eq:DMsingleflare} above we selected $\Delta T=5$ times the exponential decay timescale, i.e., the time interval during which 99\% of the flare's total energy is deposited.
\begin{figure*}[t!]
    \centering
    \includegraphics[width=.4\linewidth]{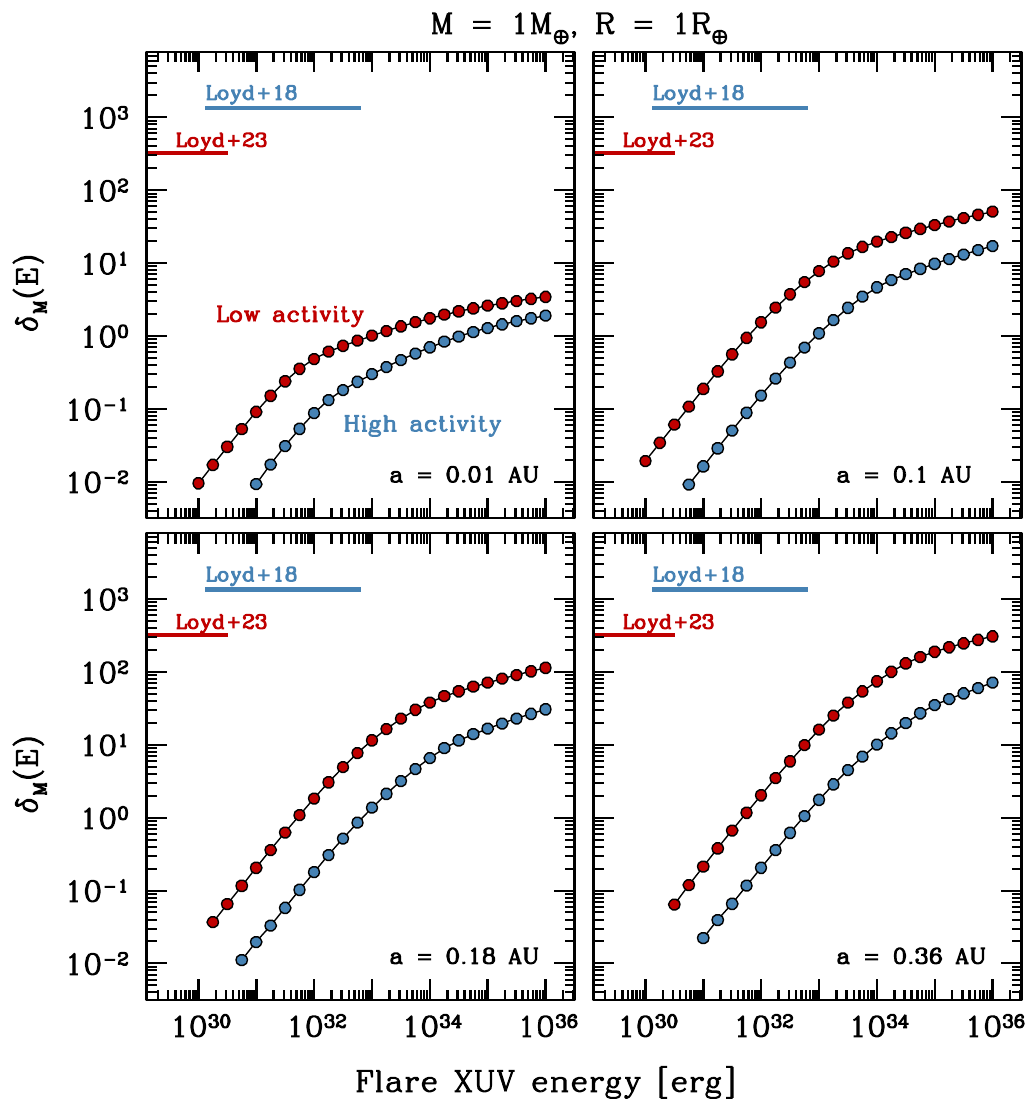}\hspace{1cm}
    \includegraphics[width=.4\linewidth]{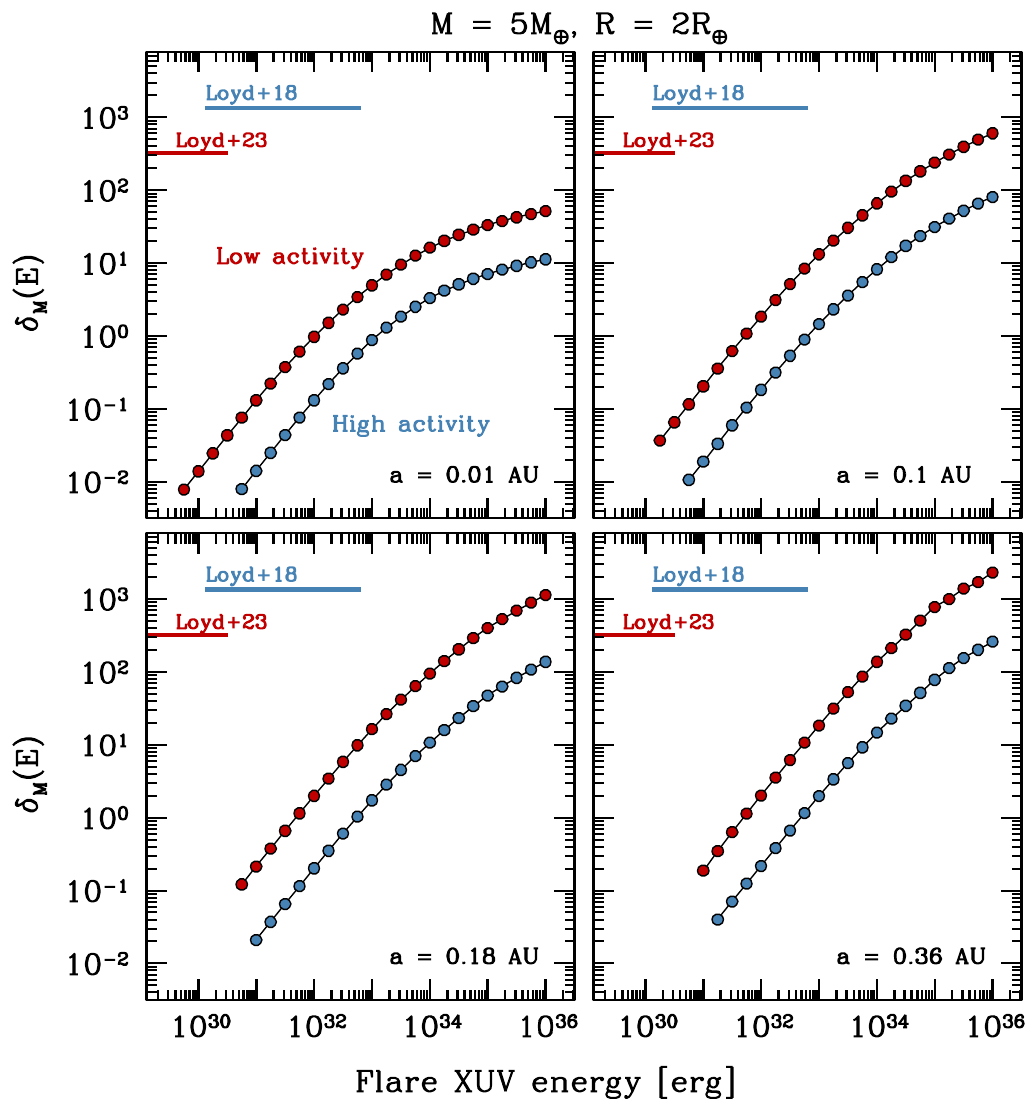}
    \caption{Fractional mass loss enhancement per flare (as given by Eq.~\ref{eq:DMsingleflare}) as a function of flare energy for both the low-activity (red) and high-activity (blue) cases, each evaluated at four different orbital distances; the 0.18-0.36 AU range brackets the extent of the HZ for the chosen host star. The horizontal lines at the top denote the energy range over which the FFDs were observationally constrained \citep{Loyd_2018, Loyd_2023}. Our simulations extend to include flare energies up to $10^{36}$ erg. The left and right panels refer to the Earth- and sub-Neptune-sized planet, respectively.}
    \label{fig:Efl_Mloss}
\end{figure*}
\subsection{Case Study: An Earth- and sub-Neptune-sized planet orbiting an early type M dwarf }
For this pilot study, we consider two exemplary systems, representative of an Earth-sized planet (with mass $M_p = 1$~\ME and radius $R_p = 1$~\RE) and a sub-Neptune-sized planet (with mass $M_p = 5$~\ME and radius $R_p = 2$~\RE), orbiting an early type M star.  
The input stellar spectrum, taken from the HAZMAT repository\footnote{\url{https://archive.stsci.edu/hlsp/hazmat}} \citep{Peacock_2019b} is that of the M2.5V type dwarf GJ 436 (radius $R_\star = 0.455~R_\odot$; mass $M_\star = 0.507~M_\odot$; effective temperature $T_{\rm eff} = 3416$~K; rotation period $P_{rot} = 44~$days), which has a well-constrained FFD \citep{Loyd_2023}.
\par
To model the impact of flares over different evolutionary times while accounting for the progressive decline in stellar rotation and magnetic activity, we consider two evolutionary phases (or stellar ages): the first covering approximately 1 Gyr, representing a young star with high-activity, and the second spanning the subsequent 4 Gyr, characterized by low stellar activity. We will refer to these two phases as ``high-activity'' and ``low-activity'', respectively.
\par
GJ 436 is believed to be at least 4 Gyr old, so we integrate the star's spectral energy distribution (SED) to estimate the quiescent stellar XUV luminosity for the low-activity phase, yielding $\LXUV^Q \simeq 10^{27.95}$ erg/s. For the star's low-activity phase, we adopt the same FFD as measured by \citet{Loyd_2023} in the FUV-130 $[1150 - 1430]$~\r{A} range for GJ 436. 

To estimate $\LXUV^Q$ over the high-activity phase, we calculate the stellar evolutionary tracks using the {\texttt Mors} code \citep{Johnstone_2021}, under the assumption that the SED shape remains constant over time. This results in a roughly tenfold increase in quiescent XUV luminosity, yielding $\LXUV^Q \simeq 10^{28.93}$ erg/s.
For the high-activity phase FFD, we use the fit to a sample of flares observed from 40 Myr old M dwarfs in the Tucana–Horologium group, and measured over same energy band \citep[see Eq. 3 in][]{Loyd_2018}.
\par
The FUV-130 energy is converted to XUV by using the ratio of the quiescent stellar flux in these two bands, as measured by HAZMAT: $F_{[100 - 912] ~\text{\r{A}}}/F_{[1150 - 1430] ~\text{\r{A}}}  \approx 6.56$. Converted to XUV, the high-activity phase differential FFD is
\begin{align}
   \frac{d \dot{N}}{dE} = 4.6094 \cdot 10^{-29}\left(\frac{E} {10^{30}}\right)^{-1.61} \; [\text{day}^{-1} 
     \text{erg}^{-1}],
    \label{eq:FFD_inactive}
\end{align}
while the low-activity FFD is 
\begin{align}
\frac{d \dot{N}}{dE} = 1.4579 \cdot 10^{-30}\left(\frac{E} {10^{30}}\right)^{-1.74} \; [\text{day}^{-1} 
 \text{erg}^{-1}].
    \label{eq:FFD_active}
\end{align}
Formally, the aforementioned FFDs are valid only within a specific range of flare energies for which flares were actually observed: approximately between $10^{27.7}$~erg and $10^{29.7}$~erg for the low-activity phase and between $10^{29.3}$~erg and $10^{32}$~erg for the high-activity phase.
Observations are likely biased against detecting both rare high-energy flares and faint low-energy flares. To account for the probable presence of both types, FFDs are typically extrapolated to cover a broader range of energies. In the following analysis, we assume that both FFDs extend between $10^{27}$~erg and $10^{36}$~erg. However, this assumption could lead to an overestimation of flare rates. The FFDs are expected to flatten at low energies and become steeper at high energies \citep[e.g.,][]{Veronig2002,Hawley2014,Silverberg2016}.
\subsection{Computational setup and modeling}
We perform time-dependent simulations using ATES\footnote{Publicly available at \url{https://github.com/AndreaCaldiroli/ATES-Code}} \citep{Caldiroli_2021}. Given the planetary system's parameters—such as planet mass and radius, orbital distance, equilibrium temperature, and stellar XUV irradiation—the code computes the temperature, density, velocity, and ionization fraction profiles of the irradiated H-He atmosphere, along with the instantaneous mass loss rate.
The photoionization equilibrium model includes cooling via bremsstrahlung, recombination, and collisional excitation and ionization for a primordial atmosphere composed entirely of atomic hydrogen and helium (with a He/H number density ratio set to 1/12 throughout this work), whilst also accounting for the advection of the different ion species.
The applicability range of ATES is well understood \citep{Caldiroli_2021}. For a sub-Neptune-sized planet, the code is expected to yield reliable steady-state solutions across the entire range of XUV irradiances considered in this work, i.e., between $10^{2}$ and $10^{5.5}$~erg~s$^{-1}$~cm$^{-2}$.

We modify the code to accommodate a time-varying input stellar XUV flux as described in the previous Section (Eq.~\ref{eq:Ltime}). In this process, we adjust only the stellar SED normalization while keeping the shape unchanged throughout each flare. 
For each planet, we run two sets of simulations for the low- and high-activity case, each with four different orbital separations, i.e., $a = 0.01$, $0.1$, $0.18$ and $0.36$~AU. The latter two correspond to the inner and outer edges of the HZ for a star like GJ~436 \citep{Kopparapu2013}. 
\begin{figure*}[t!]
    \centering
    \includegraphics[width=.55\linewidth]{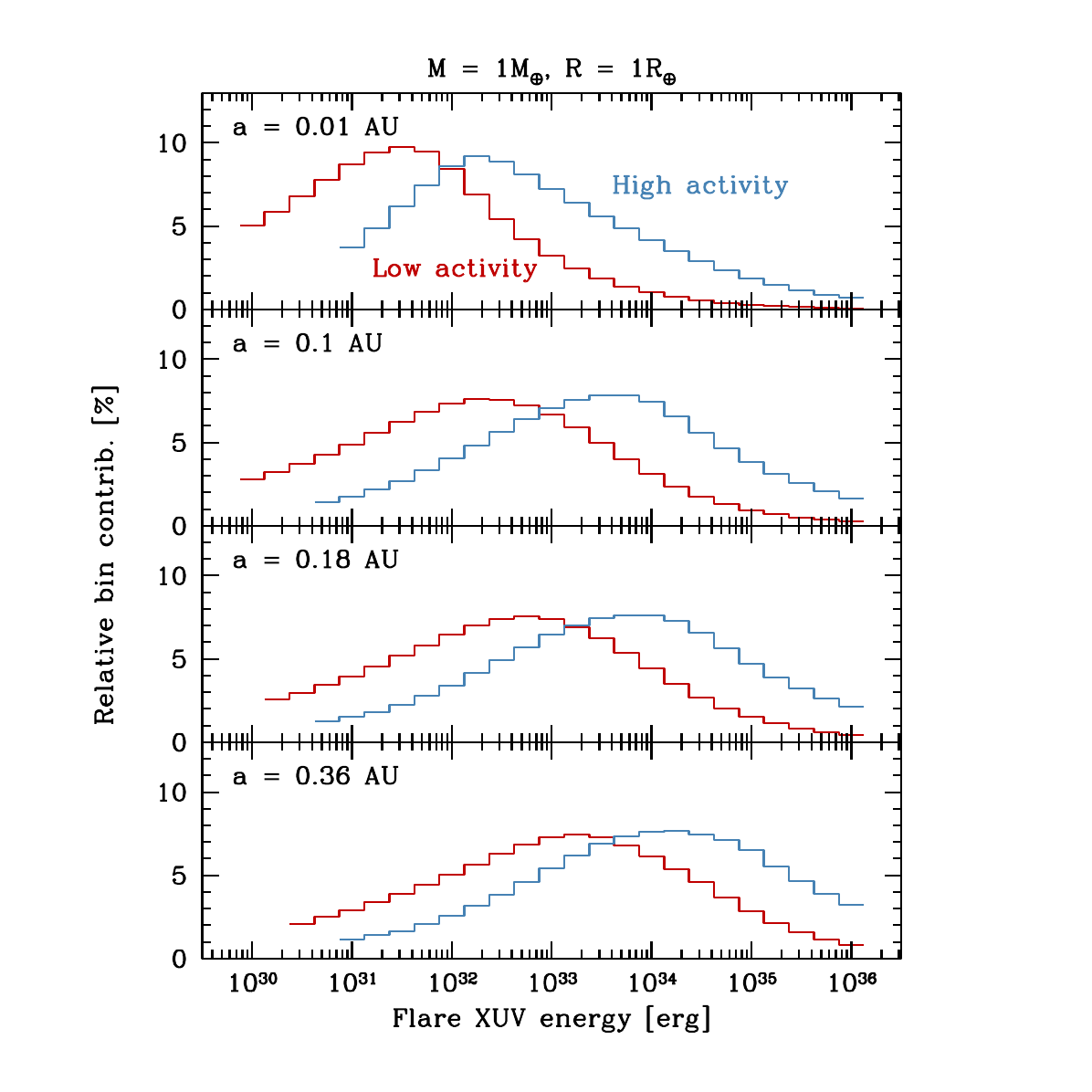}\hspace{-2cm}
    \includegraphics[width=.55\linewidth]{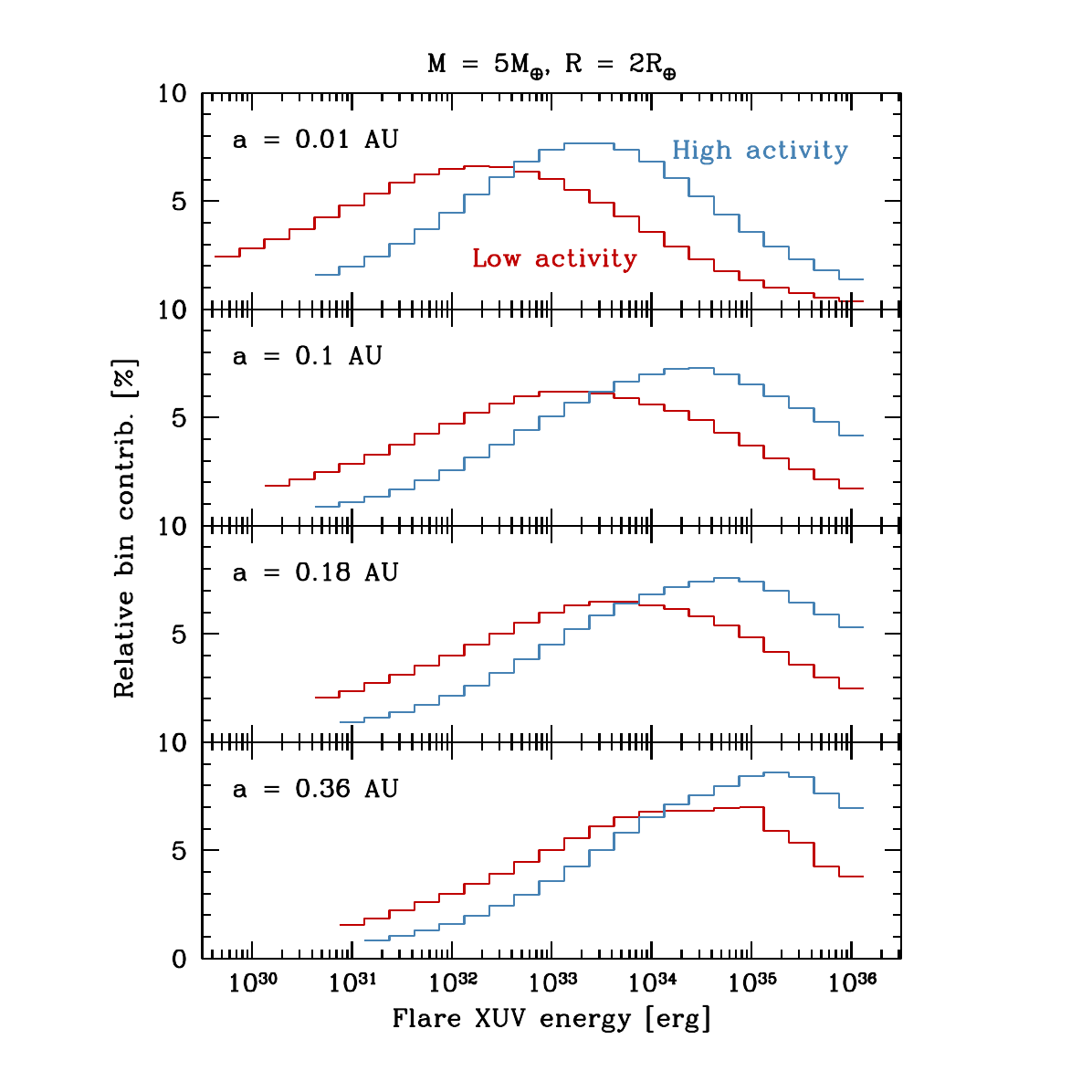}
    \caption{Fractional contributions to the flare-driven mass loss enhancement (Eq.~\ref{eq:DMflare}) from different flare energies, shown for the high-activity (blue histogram) and low-activity (red) phase, and for four orbital separations. The left and right panels correspond to the Earth-sized planet and the sub-Neptune-sized planet, respectively.}
    \label{fig:DMM_rel_integ}
\end{figure*}
For each simulation, we first obtain the stationary solution using the quiescent XUV flux as input. This solution is then used as the initial condition to simulate multiple, independent flare events with energies ranging between $10^{27-36}$~erg. 
For each flare, we monitor the time variation of the mass loss rate at the system's Hill radius, where the gas becomes gravitationally unbound from the planet. The simulation starts at $t=0$, when the flare is injected, and continues until the atmosphere has completely relaxed to the initial stationary state. We impose a minimum evolution time of $10$~hours, consistently with the typical magnitude of the sound speed and the dimension of the domain of the simulation, to ensure that all perturbations induced by the flares have time to cross the Hill radius surface. The hydrodynamic time step used by ATES for time integration is on the order of $\sim 1$ s, which allows the code to capture the input flare light curves with sufficient resolution. 

The mass contrast $\delta_M(E)$ is calculated by integrating the mass loss rate at the system Hill radius (i.e. where the gas becomes gravitationally unbound from the planet) over the duration of the simulation. Since the lowest energy flares may produce perturbations smaller than ATES' convergence tolerance (i.e., causing changes in mass loss rates of no more than 0.1\% compared to the steady state value), we exclude from the results any flares for which $\delta_M \lesssim 10^{-3}$. From the value of the mass contrast we calculate the mass loss enhancement via Eq.~\ref{eq:DMflare} and using the expression of the FFDs in Eqs~\ref{eq:FFD_inactive} and~\ref{eq:FFD_active} for the low- and high-activity phase, respectively. 
\section{Results}
\label{sec:res}
\subsection{Flare-driven mass loss enhancement}
Fig.~\ref{fig:t_Mdot} illustrates the effects induced by distinct flares of varying energies on the planet's atmosphere. These results are obtained for the low-activity scenario at an orbital distance of $a=0.1$~AU. Despite the relatively short duration of the flares, the perturbations induced on the planet's atmosphere take several hours to reach the Hill radius.

Fig.~\ref{fig:Efl_Mloss} presents the results from the complete set of simulations, showing the mass contrast $\delta_M(E)$ (calculated over an interval of 1.4 hr) plotted as a function of flare energy for both the low-activity (red symbols) and high-activity (blue symbols) phase across four different orbital distances. 
The fractional mass loss enhancements are determined according to Eq.~\ref{eq:DMflare}, and summarized in Table~\ref{tab:results}. Qualitatively, the mass loss enhancement is greater (by a factor about 10) for the high-activity phase, despite the lower absolute value of the mass contrast per flare. This is due to the $1$–$2$ orders of magnitude higher number of flares emitted per unit time. 
The $\delta_M(E)$ curves increase almost linearly at lower energy ranges and progressively flatten, approaching a $\propto E^{1/2}$ dependence at higher energy levels. We will justify this behavior analytically in Sect.~\ref{sec:discussion}. Regardless of the activity level, the energy at which the slope changes increases with orbital distance. The approximate tenfold difference in normalization between the curves corresponding to the same orbital distances is related to the approximate tenfold difference in XUV luminosities over the two stellar activity phases. Generally, more energetic flares are required to perturb the atmosphere of a planet exposed to higher levels of quiescent irradiation.
\begin{table}[!ht]
    \centering
    \caption{Summary of the values of the quiescent mass loss rate ($\dot{M}^Q$ [g s$^{-1}$]) and of the mass loss enhancement ($\Delta\MlostF/\MlostQ$) obtained from the simulations at the four orbital distances ($a = 0.01, 0.1, 0.18, 0.36$~AU) for the two activity levels (low and high) considered.}
    \renewcommand{\arraystretch}{1.5}
    \begin{tabular}{c c c | c c}
                                                  &
        \multicolumn{2}{c}{\textbf{Earth-like}}   & 
        \multicolumn{2}{c}{\textbf{Sub-Neptune}}  \\
        \cline{2-5}
                                  &  
        $\log_{10}\dot{M}^Q$      & 
        $\Delta\MlostF/\MlostQ$   &
        $\log_{10}\dot{M}^Q$      & 
        $\Delta\MlostF/\MlostQ$   \\ 
        \hline
        \hline 
        & \multicolumn{4}{c}{\textbf{a = 0.01 AU}} \\
        \hline 
        Low act.  & 10.21 &  0.93\%  & 10.62 &  2.43\%  \\ 
        High act. & 10.76 &  9.47\%  & 11.22 &  27.3\%  \\
        \hline
        & \multicolumn{4}{c}{\textbf{a = 0.1 AU}} \\
        \hline 
        Low act.  & 8.64 &  3.43\%  & 8.88 &  6.33\%  \\
        High act. & 9.47 &  35.3\%  & 9.74 &  65.9\%  \\ 
        \hline
        & \multicolumn{4}{c}{\textbf{a = 0.18 AU (Inner HZ)}} \\
        \hline 
        Low act.  & 8.12 &  4.62\%  & 8.37 &  8.07\%  \\ 
        High act. & 9.05 &  48.7\%  & 9.29 &  87.8\%  \\ 
        \hline
        & \multicolumn{4}{c}{\textbf{a = 0.36 AU (Outer HZ)}} \\
        \hline 
        Low act.  & 7.48 &  6.55\% & 7.77 &  10.9\% \\ 
        High act. & 8.50 &  74.4\% & 8.72 &  126\%  \\
        \hline
    \end{tabular}
    \label{tab:results}
\end{table}

To visualize which characteristic flare energy is responsible for most of the enhancement, Fig.~\ref{fig:DMM_rel_integ} shows the percentage contribution of each energy bin to $\Delta M^F/M^Q$ for all the simulations performed. The value of this energy depends on both the orbital separation and the activity level--and thus the quiescent XUV luminosity--of the star. For a fixed orbital separation, the characteristic flare energy which is responsible for the greatest mass loss enhancement is always higher for the high-activity phase. 

We now focus on the impact of varying the orbital distance, or equivalently, the stellar irradiance. This is particularly important for assessing the overall effect of flares on the atmospheres of planets within the HZ, as the removal of any primordial H-He envelope is thought to be a necessary condition for potential habitability. Regardless of the stellar activity level (i.e., choice of FFD), the larger the orbital separation the higher the characteristic flare energy which produces the greatest mass loss enhancement. 
As detailed in Sect.~\ref{sec:discussion}, this is due to the combined effect of the shift in the $\delta_M(E)$:$E$ curve (Fig. \ref{fig:Efl_Mloss}) and the different slopes of the FFDs. For the Earth-like planet, the fractional mass loss enhancement during the high-activity phase increases from $\approx 9.5\%$ at 0.01 AU to $\approx 35\%$ at 0.1 AU, rising to $49\%$ at 0.18 AU and $74\%$ at 0.36 AU. In comparison, the enhancement during the low-activity phase increases from approximately $1\%$ to $3.4\%$, and further to $4.6$--$6.6\%$ at 0.18--0.36~AU. A similar trend is evident for the sub-Neptune planet, where the fractional mass loss enhancement increases from $27\%$ at 0.01 AU to approximately $66\%$, $88\%$, and $126\%$ at 0.1, 0.18, and 0.36 AU, respectively, during the high-activity phase. In the low-activity phase, the enhancement rises from $2.4\%$ to $6.3\%$, and further to $8.1$--$11\%$ at 0.18--0.36~AU.
We conclude that, when the planet is closer to its host star, it is proportionally less affected by flaring activity. In other words, at 0.01~AU, atmospheric loss is driven primarily by the elevated level of quiescent irradiation rather than by flares. 

As the planet moves further away from the star, the quiescent mass loss rate $\dot{M}^Q$ decreases, and flares eventually become the dominant factor in mass loss. However, it is important to note that at sufficiently large distances, photoevaporation becomes negligible, and even the most energetic stellar flares would no longer have enough energy to drive an outflow (i.e., Jeans escape regime). The primary reason for this somewhat counterintuitive trend is the fraction of time spent in energy vs. recombination-limited mass loss regimes. At 0.01 AU, both planets are already in a recombination-limited mass loss regime even when the stellar flux is at its quiescent level; in this regime, $\dot{M} \propto F_{\rm XUV}^{0.5}$. Conversely, the larger orbital separation case is one in which mass loss is energy-limited, and $\dot{M} \propto F_{\rm XUV}$, even for relatively bright flares.
%
%---------------------------------------------------------------------------- %
%
\subsection{Evaporated mass fractions, with and without flares}

Using the results in Table~\ref{tab:results}, we can estimate the cumulative impact of stellar flares by assuming a simplified, two-phase evolution model. In this model, the star undergoes a 1 Gyr-long high-activity phase followed by a 4 Gyr-long low-activity phase \citep{West2008}.
Based on the quiescent mass loss rates from the ATES simulations, we find that, in the absence of flares, the total mass lost over 5 Gyr for the Earth-like planet is $0.65$, $0.025$, $0.009$, and $0.003 M_\Earth$ at orbital distances of 0.01, 0.1, 0.18, and 0.36 AU, respectively; for the sub-Neptune, the corresponding values are $1.75$, $0.045$, $0.015$, and $0.004 M_\Earth$.
When flares are included, the total mass lost increases to approximately $0.68$, $0.031$, $0.012$, and $0.004 M_\oplus$ for the Earth-like planet, and $2.02$, $0.066$, $0.025$, and $0.007 M_\oplus$ for the sub-Neptune, at orbital distances of 0.01, 0.1, 0.18, and 0.36~AU, respectively. These values correspond to fractional enhancements of about $5\%$, $24\%$, $33\%$, and $33\%$ for the Earth-sized planet, and $15\%$, $47\%$, $67\%$, and $75\%$ for the sub-Neptune at the same separations, respectively. We emphasize that the actual values depend on the assumed duration of each activity phase.

When compared to a more simplistic approach that assumes constant evaporation efficiency (e.g., energy-limited escape), our calculations yield systematically lower flare-driven mass loss enhancements. Using the energy-limited equation to calculate the mass loss rate, the enhancements are approximately $26\%$ and $350\%$ for the low- and high-activity phases, respectively, regardless of planet size or orbital separation. Over 5 Gyr, the total mass lost is $1.79$, $0.08$, $0.03$, and $0.008 M_\oplus$ for the Earth-like planet, and $5.07$, $0.152$, $0.053$, and $0.014M_\oplus$ for the sub-Neptune, at orbital distances of 0.01, 0.1, 0.18, and 0.36 AU, respectively. At greater distances from the star, where the XUV flux is low, the energy-limited model provides a good approximation of the escape rate. However, this is not the case at closer orbital separations. 

Our analysis shows that, overall, flares have a modest impact--less than a factor of two--on increasing the total mass lost, with the greatest cumulative enhancement occurring when the planet is in the closest orbit. At larger orbital separations, the total mass lost is much lower due to the reduced quiescent mass loss rate. However, the relative enhancement in mass loss between the flaring and non-flaring cases is greater at larger separations.
\begin{table*}[!ht]
    \centering
    \caption{Summary of the values of the total mass lost in the simplified two-phase evolution of the activity of the star.} 
    \renewcommand{\arraystretch}{1.5}
    \begin{tabular}{c c c c c | c c c c}
                                                 &  
        \multicolumn{4}{c}{\textbf{Earth-like}}  &  
        \multicolumn{4}{c}{\textbf{Sub-Neptune}}  \\
        \cline{2-9}
                                               &  
        High act.                              &  
        Low act.                               & 
        \multicolumn{2}{c|}{Total (0-5 Gyrs)}  &
        High act.                              &  
        Low act.                               & 
        \multicolumn{2}{c}{Total (0-5 Gyrs)}   \\
                      &  
        (0-1 Gyrs)    &  
        (1-5 Gyrs)    & 
        $[M_\Earth]$  &
        $[\% M_p]$    &
        (0-1 Gyrs)    &  
        (1-5 Gyrs)    & 
        $[M_\Earth]$  &
        $[\% M_p]$    \\
        \hline
        \hline 
        & \multicolumn{8}{c}{\textbf{a = 0.01 AU}}   \\
        \hline 
        w/o flares & 0.301 & 0.345 & 0.646 & 64.6\% & 0.883 & 0.872 & 1.754 & 35.1\% \\
        w/ flares  & 0.330 & 0.348 & 0.678 & 67.8\% & 1.123 & 0.893 & 2.016 & 40.3\% \\  
        \hline
        & \multicolumn{8}{c}{\textbf{a = 0.1 AU}} \\
        \hline 
        w/o flares & 0.016 & 0.009 & 0.025 & 2.5\% & 0.029 & 0.016 & 0.045 & 0.90\% \\
        w/ flares  & 0.021 & 0.010 & 0.031 & 3.1\% & 0.049 & 0.017 & 0.066 & 1.32\% \\  
        \hline
        & \multicolumn{8}{c}{\textbf{a = 0.18 AU (Inner HZ)}} \\
        \hline 
        w/o flares & 0.006 & 0.003 & 0.009 & 0.9\% & 0.010 & 0.005 & 0.015 & 0.30\% \\
        w/ flares  & 0.009 & 0.003 & 0.012 & 1.2\% & 0.020 & 0.005 & 0.025 & 0.50\% \\  
        \hline
        & \multicolumn{8}{c}{\textbf{a = 0.36 AU (Outer HZ)}} \\
        \hline 
        w/o flares & 0.002 & 0.001 & 0.003 & 0.3\% & 0.003 & 0.001 & 0.004 & 0.08\% \\
        w/ flares  & 0.003 & 0.001 & 0.004 & 0.4\% & 0.006 & 0.001 & 0.007 & 0.14\% \\  
        \hline
    \end{tabular}
    \label{tab:mloss}
\end{table*}
\begin{figure}[b!]
    \centering
    \includegraphics[width = .9\columnwidth]{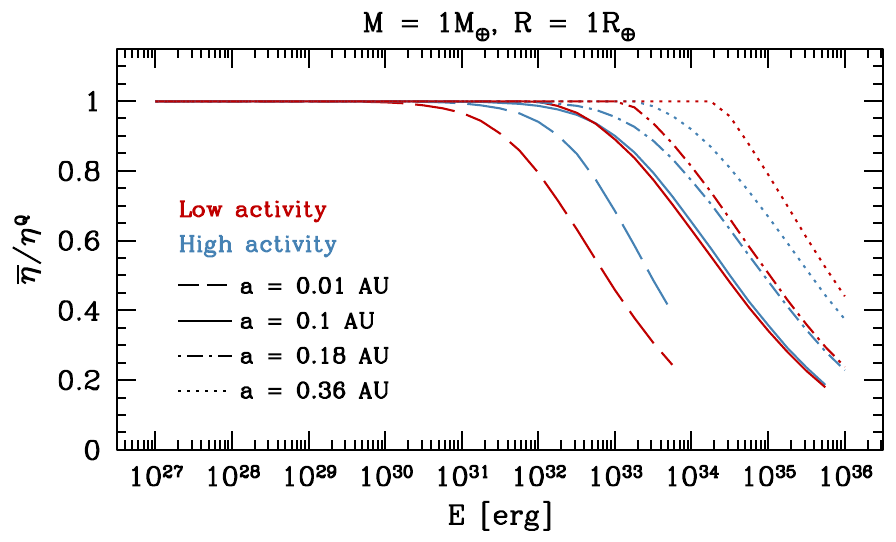} \\[10pt]
    \includegraphics[width = .9\columnwidth]{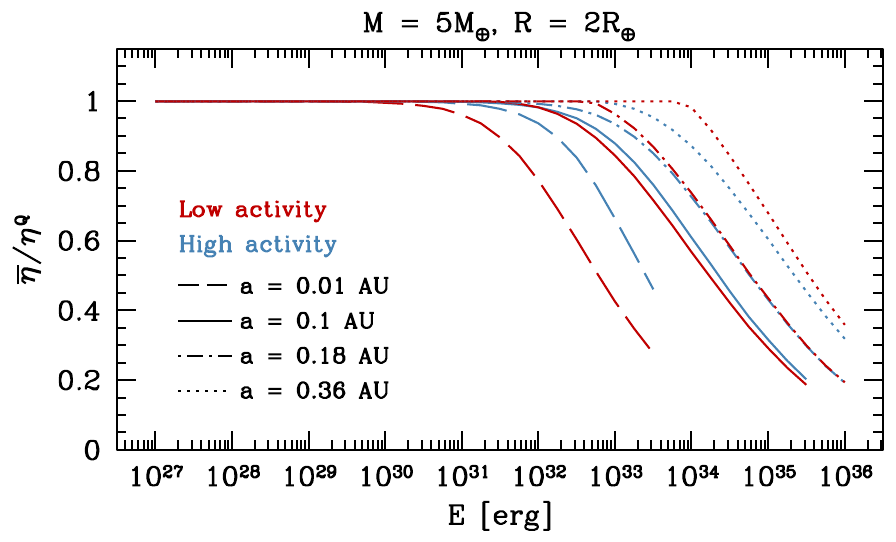}
    \caption{Ratio of the average efficiency of the outflow to the value of the quiescent efficiency as a function of the flare energy for the high-activity phase (blue) and the low-activity phase (red). Efficiencies are calculated using the analytical fitting formula given in Appendix A of \citet{Caldiroli_2022}. The top and bottom panels correspond to the Earth-sized planet and the sub-Neptune-sized planet, respectively.} 
    \label{fig:eta_ratio}
\end{figure}
%
%---------------------------------------------------------------------------- %
%
\section{Discussion}
\label{sec:discussion}
\subsection{Analytical approach}
The main results from our simulations can be summarized as follows: (i) for a given FFD, there exists a characteristic stellar flare energy which maximizes the relative contribution to the mass loss (Fig.~\ref{fig:DMM_rel_integ}); (ii) the fractional impact of flares on mass loss is larger at larger orbital distances (Table~\ref{tab:results}); (iii) regardless of planet size, stellar activity level and orbital distance, flares have a modest contribution to the cumulative mass loss (Table~\ref{tab:mloss}).
Below, we demonstrate that these somewhat counterintuitive behaviors can be understood by considering the functional dependence of the evaporation efficiency on irradiance, in conjunction with the specific shape of the FFDs.
\par
Without loss of generality, the total atmospheric mass lost within a given time interval can be expressed as $\Delta M \approx \eta\, 3\,{\cal F}_{\text{XUV}}/(4GK\rho)$, where $\eta$ is the evaporation efficiency, ${\cal F}_{\text{XUV}}$ is the stellar XUV fluence at the planet's surface, $G$ is the gravitational constant, $K$ accounts for the gravitational potential of the star, and $\rho$ is the planetary mass density \citep[see, e.g.,][]{Earkev_2007, Sanz-Forcada2011}.
In energy-limited regime, a large fraction of the absorbed stellar energy is converted into heat and adiabatic expansion, implying $ \Delta M \propto \fxuv$, with a nearly constant value for $\eta$ \citep[typically between 0.1 and 1; see, e.g.,][]{Caldiroli_2022,Salz2016b}. This approximation is no longer valid at high irradiation \citep[$\fxuv \gtrsim 10^4$~erg~s$^{-1}$cm$^{-2}$; see, e.g.,][]{MurrayClay_2009}, as cooling via radiative recombinations (chiefly Ly$\alpha$) becomes very efficient, at the expense of expansion work. In this recombination-limited regime, the evaporation efficiency becomes flux-dependent and decreases with increasing irradiation according to $\eta \propto \fxuv^{-1/2}$ \citep{MurrayClay_2009, Owen_2016}. 
We emphasize that planetary gravity is the primary factor in determining the efficiency of atmospheric evaporation \citep{Salz2016b}. While Earth-sized and sub-Neptune-sized planets can transition from energy-limited to recombination-limited regimes, gas giants can never be energy-limited. The reader is referred to \citet{Caldiroli_2022}, and references therein, for an in-depth discussion of the dependence of $\eta$ on $\fxuv$. \par
\par
Going back to the impact of a single flare of energy $E$ and duration $\Delta T$, the resulting mass contrast can be written as: 
\begin{align}
    \delta_M(E) = \dfrac{1}{\Delta T} \int_{\Delta T} { \frac{\eta(t)}{\eta^Q} \frac{\fxuv(t)}{F^Q_{\rm XUV}}\, dt} - 1. 
    \label{eq:DMM_theo}
\end{align} 
We approximate the time-dependent flaring flux $\fxuv(t)$ with its value averaged over the flare duration, i.e.,  
$\bar{F}_{\rm XUV} = \left( 1 + E/E^Q\right)F^Q_{\rm XUV}$, where $E^Q = \Delta T\cdot\LNF$ is the energy emitted by the quiescent star in the same time interval $\Delta T$.
We now define an average flare efficiency as: $\bar{\eta}\equiv  \eta(\bar{F}_{\rm XUV})$, i.e., the efficiency at average flux. 
With these definitions the expression for the mass contrast becomes:
\begin{align}
    \delta_M(E) = \dfrac{\bar{\eta}}{\eta^Q}\left( 1 + \frac{E}{E^Q}\right)  - 1. 
    \label{eq:DMM_theo_avg}
\end{align} 
For flare energies that are low compared to $E^Q$, the evaporation efficiency remains unchanged from the quiescent state, implying that $\delta_M \propto E$. This is true regardless of the evaporation regime, whether it is energy-limited or recombination-limited. Conversely, if $E \gg E^Q$, the outflow is most likely recombination-limited, which implies $\delta_M \propto E^{1/2}$, since $\bar{\eta} \propto \bar{F}_{\rm XUV}^{-1/2}$.  
\par
As shown in Fig.~\ref{fig:eta_ratio}, the transition energy between the $\delta_M \propto E$ and $\delta_M \propto E^{1/2}$ regimes corresponds to the flare energy value where the ratio $\bar{\eta}/\eta^Q$ begins to differ significantly from unity. Different FFDs and orbital distances result in different transition energies. However, it is important to note that the flare energy at which this transition occurs does not correspond to the flare energy that maximizes the flare's contribution to mass loss. To understand the latter, we must examine the interplay between the dependence of $\delta_M$ on flare energy $E$ and the functional shape(s) of the FFD, as follows. 

To encompass the full range of behavior between energy- and recombination-limited escape, we express the mass contrast as a generic power-law function of energy: $\delta_M(E) \propto E^\beta$, where $\beta$ varies between 0.5-1. Next, we consider the product $E\,\delta_M\,d\dot{N}/dE$; a maximum in this quantity corresponds to the characteristic flare energy that maximizes the mass-loss enhancement due to flares (see Eq.~\ref{eq:DMflare} and Eq.~\ref{eq:DMsingleflare}).
The differential distribution of the number of flares $d\dot{N}/dE$ scales as $\propto E^{-\alpha}$, where $\alpha=1.74\,(1.61)$ for low-activity (high-activity) stars.
Combing the two expressions above yields $E\,\delta_M\,d\dot{N}/dE \propto E^{\beta+1-\alpha}$, which has a maximum for $\beta=0.74$ in the case of low-activity stars, and for $\beta = 0.61$ in the case high-activity stars. 
In practice, as $\beta$ decreases from 1 to approximately 0.5 as the flare energy increases, the energy-dependent ratio $\Delta M^F/M^Q(E)$ must exhibit a maximum somewhere between $E_{\rm min}$ and $E_{\rm max}$. Critically, this means that neither rare super-flares nor frequent low-amplitude flares have a substantial impact on long-term mass loss enhancements due to flares.
\begin{figure*}
    \centering
    \sidecaption
    \includegraphics[width=12cm]{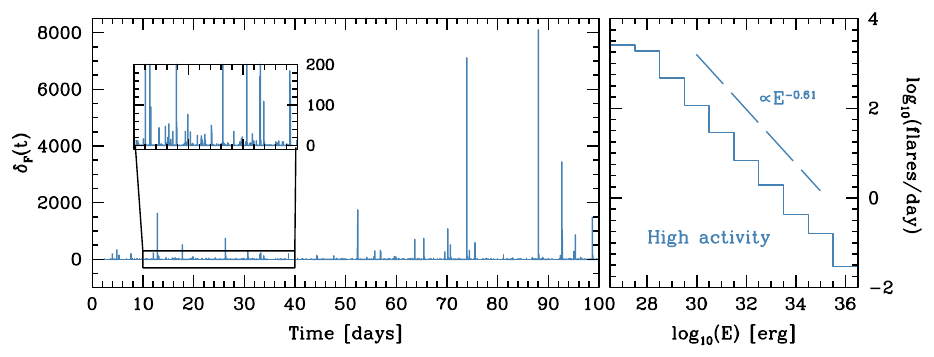}
    \caption{The left-hand panels display one-year light curves corresponding to the high-activity stellar phase. The right-hand panel presents the flare occurrence rate as a function of energy over the same time interval.}
    \label{fig:series}
\end{figure*}
\subsection{Simulations with realistic flare time series}
\label{subsec:timeseries}
The results presented thus far are based on the implicit assumption that atmospheric relaxation times in response to a flare are shorter than the average interval between distinct flares. This implies that the outflowing atmosphere reaches a steady state before the onset of the next flare. However, our simulations indicate that the atmosphere may take several hours to relax following the most energetic flares (see Fig.~\ref{fig:t_Mdot}). Consequently, the assumption that flares can be treated independently may not always hold. 

This is especially relevant in the high-activity case, where the flaring rate is elevated. For a star in the low-activity regime, the average time interval between low-energy flares (i.e., $E \approx 10^{30-31}$erg; see Fig.\ref{fig:DMM_rel_integ}) is approximately $10$--$30$~hours. This interval is of the same order as the atmospheric relaxation time, which is about $5$ hours at an orbital separation of $0.01$ AU and approximately $40$ hours at $0.36$ AU. Thus, for a star in this activity regime, it is reasonable to assume that flares can be treated independently. In the following, we focus on the high-activity phase for the sub-Neptune, for which the cumulative mass loss is expected to be much higher compared to the Earth-like planet. 

To test a more realistic scenario, we construct a template flare time series by sequentially appending multiple flares. Each flare energy is randomly selected from the input FFD, and the flare contrast is constructed as described in Sect.~\ref{sec:method}. The total duration of the time series is set to 100 days, ensuring that the distribution of flare energies is adequately sampled over the entire range from $10^{27}$ to $10^{36}$ erg. The resulting time series is illustrated in Fig.~\ref{fig:series}.
We note that for this procedure to be robust, it is necessary to conduct multiple simulations with different random realizations of the time series.

Using 10 realizations of the high-activity time series described above as inputs, we repeat the ATES simulations described in the previous Sections. The total mass loss enhancement is calculated by integrating the mass loss rate at the Hill radius over the entire duration of the simulation(s). 

We find fractional mass loss enhancements of $27.61\pm 2.0\%$, $64.6 \pm 6.6 \%$, $82.6 \pm 8.5\%$ and $123.7 \pm 15.2\%$ at orbital distances of $0.01$, $0.1$, $0.18$ and $0.36$ AU, respectively. These values translate to a cumulative mass loss over 5 Gyr of $1.998 \pm 0.018~M_\Earth$, $0.064 \pm 0.002~M_\Earth$, $0.024 \pm 0.0009~M_\Earth$ and $0.008 \pm 0.0004~M_\Earth$ at the four orbital distances. The errors are calculated as the standard deviation of the values obtained from the different realizations. 
The time-series results are consistent with those reported in Table\ref{tab:results} and Table\ref{tab:mloss} within 1-2$\sigma$. 

We conclude that assuming independent flares is a valid approximation for calculating the integrated contribution of flares to mass loss enhancement. This is justified by considering the average time interval between flares of the characteristic energy responsible for the greatest enhancement. From Fig.\ref{fig:DMM_rel_integ}, we find that the peak energy is approximately $10^{33.5}$ erg at $0.01$ AU and $10^{35.5}$erg at $0.36$ AU. These values correspond to average intervals of roughly 20 hours and 10 days, respectively, both of which exceed the corresponding atmospheric relaxation times, which range from about 5 to 40 hours.
\subsection{Summary and conclusions}

This work represents the first attempt to model the effect of stellar flares on atmospheric evaporation—specifically regarding the H-He envelope—using a time-dependent approach. Our results are qualitatively consistent with the conclusions of \citet{doAmaral2025}, who focused on the young M1 star AU~Mic and its planetary system, composed of one Earth-sized and two Neptune-sized planets.
In their study, the effect of flares is modeled by increasing the quiescent luminosity by an amount corresponding to the average enhancement induced by flares, with different flare FFDs adopted for various evolutionary epochs. As confirmed by our analysis, they conclude that the effects of flaring are more significant at larger orbital separations. Our approach provides physical insight into the underlying reasons for this behavior.

The primary reason for this trend is the fraction of time spent in energy vs. recombination-limited mass loss regimes. At close orbital separations, the planet is in a recombination-limited mass loss regime, regardless of whether the star is flaring or not; in this regime, the mass loss rate scales as the square root of the irradiance. The larger orbital separation case is one in which mass loss is energy-limited, and the mass loss rate scales linearly with irradiance, even for relatively bright flares. This implies that the large separation case is associated with a larger increase is $\dot M$. 

Importantly, we demonstrate the existence of a characteristic flare energy that maximizes the fractional contribution to flare-driven mass loss. Even though the actual values of this characteristic energy depend on the orbital distance and the assumed FFD shape, this ensures that neither rare super-flares nor frequent low-amplitude flares have a significant impact on the mass loss enhancements caused by stellar flares.

We conclude by emphasizing that the quantitative estimates presented in Tables 1 and 2 are specific to the chosen planetary systems and orbital distances. 
Overall, we expect that our results will qualitatively apply to other planetary systems--provided they are not gas giants--so long as they experience irradiation levels comparable to those considered in this study.

Quantitatively, the absolute values of $\dot M$ are systematically lower for the Earth-sized planets compared to those of the sub-Neptune, owing to the higher average density of a terrestrial planet. Assuming an atmospheric mass fraction of $10^{-3}$\,$M_p$ after the "boil-off" phase following disk dispersal \citep[see, e.g.,][]{Owen2016,Misener2021,doAmaral2025}, we find that, neglecting flares, an Earth-like planet at 0.18~AU (0.36~AU) would lose its H/He envelope entirely within 168 (600)Myr. When considering the effects of flares, these timescales are reduced to 113 (343)Myr, respectively.

Further variations are likely to arise from (i) evolving the planet's radius in response to mass loss, and (ii) relaxing the assumption that the stellar SED does not vary during flares. These aspects, along with considering an appropriate grid of planetary radii, masses, and orbital distances, will be the subject of future work.

\begin{acknowledgements}
We are grateful to Evgenya Shkolnik and Laura do Amaral for their valuable feedback on an earlier version of this manuscript. We also thank the Editor, Emmanuel Lellouch, for prompting us to extend our original investigation to include Earth-like planets. This research was supported in part from the Michigan Institute for Research in Astrophysics (MIRA). RS acknowledges the support of the ARIEL ASI/INAF agreement no. 2021-5-HH.0 and the support from the European Union – Next Generation EU through the grant n. 2022J7ZFRA – Exo-planetary Cloudy Atmospheres and Stellar High energy (Exo-CASH) funded by MUR – PRIN 2022.
\end{acknowledgements}
%
%------------- REFERENCES -------------
%
\bibliographystyle{aa} 
\bibliography{bibliography} 
\end{document}